\documentstyle[11pt,aaspp4]{article}
\slugcomment{Submitted to ApJ}

\lefthead{Bautista and Torres}
\righthead{A Method to Estimate $P(k)$}

\begin{document}
\def\Qrms{$Q_{\rm{rms-PS}}$}

\title{A Method to Estimate the Primordial Power Spectrum from CMB Data}

\author{Rafael Bautista\altaffilmark{1}}
\affil{Departamento de F\'{\i}sica, Universidad de los Andes,
    A.A. 4976, Bogot\'{a}, Colombia}

\and

\author{Sergio Torres\altaffilmark{2}}
\affil{Observatorio Astron\'{o}mico, Universidad Nacional,
        A.A. 2584, Bogot\'{a}, Colombia  \\
and, Centro Internacional de F\'{\i}sica, A.A. 49490, Bogot\'{a}, Colombia}
\altaffiltext{1}{e-mail: rbautist@cdcnet.uniandes.edu.co} 
\altaffiltext{2}{e-mail: storres@cdcnet.uniandes.edu.co} 

\begin{abstract}
A precise determination of the 
primordial spectrum of matter 
density fluctuations at super-horizon scales
is essential in understanding large scale
structure in the universe. Attempts to 
constrain or obtain the primordial 
spectrum using data on cosmic microwave background (CMB)
anisotropies has relied on statistical and 
correlation analyses that assume a power-law spectrum. We propose
a method to derive $P(k)$ directly from 
the CMB angular power spectrum which does
not presuppose the need to know anything about 
its functional form. The method consists of 
a direct inversion of the Sachs-Wolfe formula.
Using this new analysis technique and 
COBE data we obtain an empirical 
$P(k)$ which 1) supports a power-law parameterization 
and 2) has an amplitude and 
spectral index consistent with previous 
analyses of the same data. We obtained
for the spectral index, 
$n=1.52 \pm 0.4$ when the 2nd year COBE data
is used and
$n=1.22 \pm 0.3$ using the 4 year data set.

\end{abstract}


\section{Introduction}

The mechanism for large scale structure formation 
in the universe calls for primordial 
density fluctuations (PDF) in the early universe.
A knowledge of the spectrum
of PDF, $P(k)$, would allow to 
compute the {\it rms} mass fluctuation on a given 
scale, $\delta M/M$ and the peculiar velocity field.
Inflation predicts a scale invariant spectrum
$P(k) \propto k$ (\cite{har70,zel72}).

The availability of cosmic microwave background (CMB) 
data at large angle scales 
[COBE (\cite{ben96,ben94}), 
TENERIFE (\cite{han94}), and 
FIRS (\cite{gan93})]
has made it possible at least in principle to probe the shape of the
PDF spectrum. Theoretical uncertainties (i.e. `cosmic variance') and
experimental constraints such as sampling variance,
low signal-to-noise ratio 
and 
galactic contamination, however, impose very stringent limitations
in the ability to obtain the original spectrum. 
In order to 
deal with the effects of an equatorial cut in the 
portion of the celestial sphere dominated by diffuse galactic 
emission, \cite{gor94} have found a 
new orthogonal set of basis functions to represent
the scalar radiation field. An alternative used 
by \cite{wri94a} (hereafter WRI94) applies weights to the 
spherical harmonic decomposition of $\Delta T/T$
in order to correct for aliasing among different 
$\ell$-terms that result in the cut sphere when the 
monopole and dipole terms are removed.

Most of the analyses of CMB data aimed at 
probing $P(k)$, however, 
have been
done using a statistical 
maximum likelihood analysis on the 
angular power spectrum or
the auto-correlation function under the 
assumption of a power law for $P(k)$.
In view of the above mentioned
intrinsic and instrumental 
limitations,
the need for new and alternative analysis methods
is well justified. We propose a new technique
to obtain $P(k)$ directly from the CMB
angular power spectrum which {\em does not
assume any particular form for $P(k)$}, thus
allowing to test for deviations from power law
models. The 
method is based on a direct integration of the 
Sachs-Wolfe formula (\cite{sac67}) 
for the angular spectrum coefficients.
It is shown that by a straightforward application 
of the mean-value theorem the Sachs-Wolfe integral 
can be inverted resulting in a robust
estimate of $P(k)$ over the wave-length range
available to CMB experiments.

\section{The Algorithm}

Expressing the CMB temperature anisotropies in 
the usual spherical harmonics expansion allows one 
to define the angular power spectrum, 
$C_{\ell}$ (\cite{bon87}):

\begin{equation}
\frac{\Delta T}{T} = \sum_{\ell = 2}^{\infty} \sum_{m = - \ell}^{\ell} 
a_{\ell m} Y_{\ell m}(\theta,\phi),
\end{equation}
with
\begin{equation}
C_{\ell} \equiv \langle  |a_{\ell m}|^2  \rangle.
\end{equation}

The $C_{\ell}$'s used here are related to the 
rotationally invariant {\it rms} multipole moments used
by COBE by $\Delta T_{\ell}^2 = (2 \ell + 1) T_0^2 C_{\ell}/(4 \pi)$,
with $T_0$ the monopole temperature.
Experiments provide estimates for $C_{\ell}$ and their connection with 
theory is established by means of the 
Sachs-Wolfe effect. For large angle scales the $C_{\ell}$'s 
are (\cite{fab87}):
\begin{equation}
C_{\ell} = \left(  \frac{4 \pi}{5}  \right)^2 
\int_0^{\infty} P(k) \left(    \frac{j_{\ell}(k)}{k} \right)^2  dk,  \label{eq:sachsw}
\end{equation}
where we have made $k$ adimensional $(k \leftarrow 2ck/H_0)$ and the 
formula is valid for $\Omega_0 = 1$.

We attempt to get straightforward information about the spectrum of PDF
by a direct inversion of equation~(\ref{eq:sachsw}). 
The primary data are the angular correlation coefficients 
measured by experiments. 
For this purpose we have devised a method which, 
in addition to its simplicity,
has two basic virtues: 
1) It makes a minimum of assumptions about the form of the spectrum and 
2) it can be readily extended to include any new 
information on the $C_{\ell}$'s that might come from future experiments.

Let us introduce the integral
\begin{equation}
I_{\ell}=
\int_0^{\infty} \left(\frac{j_{\ell}(k)}{k} \right)^2 dk \, ,
\label{eq:kernel}
\end{equation} 
for $\ell \geq 2$ .

Notice that every factor in the integrand 
of~(\ref{eq:sachsw}) is either positive definite or at least non-negative
and smooth.  
This allows us to establish the following identity:
\begin{eqnarray}
P(\bar{k}_{\ell})  &  =  &
\frac{1}{I_{\ell}} \int_0^{\infty} 
P(k)\left(\frac{j_{\ell}(k)}{k} \right)^2 dk  \nonumber  \\
& = & \frac{C_{\ell}}{\left( \frac{4 \pi}{5} \right)^2 I_{\ell}}
\label{eq:invert}
\end{eqnarray}

which is an application of the mean value theorem
(\cite{str81}).  
Equation (\ref{eq:invert}) 
simply states that, 
for some value of its argument, 
here denoted as $\bar{k}_{\ell}$, the value of $P(k)$ has to match 
the right-hand-side of~(\ref{eq:invert}). 
This is true under the conditions stated above. 
Then the evaluation of~(\ref{eq:kernel}) and the 
knowledge of the $C_{\ell}$'s for a given 
set of values of $\ell$, yield the value of the left-hand-side of~(\ref{eq:invert}).
In principle it can be seen that the Sachs-Wolfe formula 
combines all scales in $k$-space in the coefficients $C_{\ell}$.
The reason why the direct inversion prescription works can be 
seen by examining the form of the kernel, $K_{\ell}(k) \equiv (j_{\ell}(k)/k)^2$, 
in equation~(\ref{eq:invert}).
The spherical Bessel functions are quasiperiodic with 
decreasing amplitude for large $k$, thus the 
kernel for each $\ell$ is also a periodic 
function but with its first peak being the 
only dominant contribution (see Fig. 1). 
The maximum 
of the peak is roughly located at $k \approx 0.7 + 1.1 \ell$
(with the dimensionless $k$ used here) thus
for each $\ell$  in the kernel one is 
probing a well defined and independent
scale. 

\placefigure{fig1}

Now we need to estimate with enough accuracy the arguments $\bar{k}_{\ell}$, 
in order to complete the table $P(k)$ vs. $k$.  
A way in which this can be accomplished 
is by recycling, under a new light, Zeldovich's recourse of 
estimating functions through the use of power laws, whenever the range 
of values of $k$ of physical interest is small 
(only very large scales in our case). Therefore, 
appealing to the same kind of reasoning, 
we proceed to obtain the values of $k_{\ell}$ by means 
the estimative relation
\begin{equation}
(\bar{k}_{\ell})^s 
\approx
\frac{1}{I_{\ell}}\int_0^{\infty} 
k^s \left(\frac{j_{\ell}(k)}{k} \right)^2 dk \, ,
\label{eq:estimator}
\end{equation}
where $s$ is any `reasonable' value that must be within a 
certain range that satisfies the criteria of convergence and physical plausibility.  
Of course, the values $\bar{k}_{\ell}$ obtained in this 
fashion can not be very sensitive to the particular $s$ chosen as estimator 
in formula~(\ref{eq:estimator}).  
In fact, the use of $k^s$ as an estimator function is not necessary, 
it is only one of the simplest that will do the job.  
Other, more elaborate, estimators could be used, but, as it will be seen, 
this is not necessary.

We have evaluated equation~(\ref{eq:estimator}) 
for the range $0.5 \leq s \leq 1.5$ and $2 \leq \ell \leq 12$, 
with the results that are shown in Table 1.  
As it can be quickly noticed, the maximum effect caused by 
varying  $s$ occurs for small $\ell$ with 
a spread around 
$\pm 7$\% in the worst case.
For larger $\ell$ these variations grow smaller and their effect 
is not of importance. 
\placetable{tbl-1}

\section{Analysis and Conclusions}

Using the Hauser-Peebles angular power estimator 
WRI94 obtain values for the $T^2_{\ell}$ coefficients
which are a linear combination of the 
$C_{\ell}$'s (see Table~1 in WRI94).
A numerical integration of the inversion formula
with $s=1.0$
and the angular power spectrum in the 
range $3 \leq \ell \leq 18$  derived from WRI94
was used to obtain the  $P(k)$ points illustrated in Fig. 2.   
The error bars come from the uncertainties in the COBE data for 
the $C_{\ell}$'s.  Coefficients beyond $\ell = 18$ were not 
included because
the angular power spectrum 
seen by COBE for those $\ell$'s is dominated by noise
and the associated angular scales are beyond COBE's 
angular resolution.
The observed angular power spectrum $C_{\ell}^{obs}$  is related
to the theoretical spectrum by 
$C_{\ell}^{obs} = G^2_{\ell} C_{\ell}$,
where $G_{\ell}$ is the beam profile in terms of the 
coefficients in an expansion in Legendre polynomials.
We have used the $G_{\ell}'s$ in \cite{wri94b}. 
The $C_{\ell}$'s are obtained by inverting 
the $T_{\ell,\ell^{\prime}}$ matrix in Table~1 of WRI94.
This is a matrix of dimension $\ell_{\rm max} \times \infty$
which formally does not have an inverse. However, for the particular 
case under consideration an approximate inverse can 
be computed by noticing that the non-zero matrix elements 
away from the diagonal, follow a scaling law, 
$T_{ij} \propto \left| i-j \right|^{-2.12}$.
The expansion of the $C_\ell$'s in terms of the 
$T_\ell$'s is truncated at a point where additional
terms contribute a negligible amount  relative to the 
measurement errors. Fig.~3 shows the resulting 
$C_\ell$'s.

One can test the power law `anzats' by attempting to fit
$P(k)$ to a function of the form $P(k) \propto Q^2 k^n$.
Here $Q$ denotes the {\it rms} quadrupole normalization,
more commonly written as $Q_{\rm rms-ps}$.
A maximum likelihood method taking into account the full 
covariance matrix was used. To find 
the model dependent covariance matrix, $M(Q,n)$,
a Monte Carlo procedure was followed:
first, the model parameters $n$ and $Q$ are fixed 
to generate realizations of the CMB angular power 
spectrum. These realizations of $C_\ell$ coefficients 
follow a $\chi^2$ distribution with $2\ell + 1$ degrees 
of freedom and have mean values given by formula (4.18) of 
Bond \& Efstathiou (\cite{bon87}). For each $C_\ell$
realization our inversion method delivers a corresponding $P(k)$. 
The average of these $P(k)$'s over the ensemble 
of $N_r$ realizations, $\langle P(k) \rangle$,  
was computed and the covariance matrix as well:
\begin{equation}
M_{ij}(Q,n) = \frac{1}{N_r-1} \sum_m^{N_r} 
(P_m(k_i) - \langle P(k_i) \rangle)
(P_m(k_j) - \langle P(k_j) \rangle) \, .
\end{equation}
Finally, the likelihood function $L(Q,n)$ was computed: 
\begin{equation}
-2 \ln L(Q,n) = d^T M^{-1}(Q,n) d + \ln det(M(Q,n)) + {\rm const}\, ,
\label{eq:cov}
\end{equation}
where the deviation vector $d$ is the difference
of a data point and the corresponding theoretical 
mean value from Monte Carlo realizations,
$d_i = P(k_i) - \langle P(k_i) \rangle$.
The covariance matrix is normalized so that 
the second term of equation (\ref{eq:cov})
equals the $\chi^2$ statistic.
This procedure was repeated for several values 
of $Q$ and $n$ forming a discrete sampling 
of $L(Q,n)$ inside a grid defined by the 
ranges $n$: $0.8 - 2.3$ in steps of $\Delta n = 0.05$ and 
$Q$: $4 - 28$ $\mu$K in steps of $\Delta Q = 0.5$ $\mu$K. 
A much finer resolution in $Q$ and $n$ was later 
obtained by two dimensional interpolation of the 
above defined grid of $L(Q,n)$ points. It was 
verified that with 5000 realizations
the results converged to a stable value. 

The bias and the errors on the estimated model
parameters were obtained using the Monte 
Carlo procedure described above but with input
synthetic data (for a fixed model) 
with known $Q_{\rm in}$ and $n_{\rm in}$.
For each input realization one obtains a set of values
$Q_{\rm max}$, $n_{\rm max}$ that maximizes the likelihood.
The mean of the
$Q_{\rm max}$, $n_{\rm max}$ points gives the bias and
their dispersion gives the actual errors.
The $n$ parameter is biased upward by $\approx 0.03$
and $Q$ is biased in the opposite direction by $\approx 0.26$.
The 1-$\sigma$ errors are $\delta n = 0.2$ and $\delta Q = 3.0$ $\mu$K.
This would give us the uncertainty due to `cosmic variance' alone. 
The error on the parameters 
due to instrumental noise was estimated and added in 
quadrature. The latter contribution to the error
was computed following the Monte Carlo procedure
explained above but instead of generating the 
model dependent $C_\ell$'s we took one single realization 
of $C_\ell$'s (which was fixed throughout the procedure)
and to it we added realizations of instrumental noise
power spectrum.
The noise coefficients $C_{\ell,{\rm noise}}$
are directly obtained from the harmonic coefficients of
$A-B$ map combinations. Since $A-B$ noise maps 
do not require galactic cut, a straightforward 
harmonic fit is applicable. 

The debiased results for which $L(Q,n)$ is 
maximum are $n = 1.52 \pm 0.4$ and 
$Q = 16.3 \pm 6.0$ $\mu$K which are 
consistent with WRI94. 
Fig. 3 shows 
the angular power spectrum corresponding to this best
fit $P(k)$ and COBE's data points.
To give an idea of the goodness of fit, 
the $\chi^2/$DOF at $L_{\rm max}$ is $30.612/14$.

\placefigure{fig2}

\placefigure{fig3}

We repeated the analysis with the
$C_\ell$ coefficients ($3 \leq \ell \leq 18$) from the 
4 year COBE data given by 
Tegmark (\cite{max96}). For these data,
the maximum 
likelihood analysis gives 
$n = 1.22 \pm 0.3$ and 
$Q = 16.3 \pm 4.5$. 
The analysis also 
reveals that these 
parameters are anticorrelated. 
That is, values of $Q$ and $n$
that follow the relation 
$Q(n) = 19.9 \exp[0.756(1-n)]$
lay approximately inside the 2-$\sigma$
contour level.

Our results for $Q$ and $n$ are consistent
(within the error bars)
with those obtained by the COBE group which 
are summarized in Table 4 of WRI94 and 
in Table 2 of Bennett et al. (\cite{ben96})
and depending on the analysis method or the 
way the data was prepared (i.e. which map
combination, exclusion or not of the quadrupole term, 
beam shape filter, etc) their results for $n$ range 
from $1.02 \pm 0.4$ to $1.42^{+0.49}_{-0.55}$
for the 2 year results and 
from $1.23^{+0.23}_{-0.29}$ to $1.30^{+0.30}_{-0.34}$
for the 4 year results. 
One important fact worth
noticing is that independent of the $n$ values, 
a power law form for the spectrum of PDF is indeed 
consistent with the $P(k)$ obtained here directly 
from the CMB data without an {\em a priori} assumption 
about its shape. 

\acknowledgments

We thank E. L. Wright for providing
COBE's angular power spectrum.
E. Mart\'{\i}nez-Gonz\'{a}lez 
and the anonymous referee gave us very useful comments.
S.T. was funded by COLCIENCIAS and CINDEC-Universidad Nacional 
de Colombia.

\clearpage
\begin{table*}
\begin{center}

\begin{tabular}{rrrr}

$\ell$  & 
\multicolumn{1}{c}{$\bar{k}_{\ell}(s=1.0)$\tablenotemark{a}} & 
$\Delta \bar{k}_{\ell}(s=1.5) $ &   
$\Delta \bar{k}_{\ell}(s=0.5) $ \\
\tableline

2   & $  2.782 $ & $ +0.18 $ & $ -	0.16  $ \\
3   & $  4.168 $ & $ +0.20 $ & $ -	0.17  $ \\
4   & $  5.494 $ & $ +0.23 $ & $ -	0.18  $ \\
5   & $  6.790 $ & $ +0.26 $ & $ -	0.20  $ \\
6   & $  8.067 $ & $ +0.28 $ & $ -	0.22  $ \\
7   & $  9.329 $ & $ +0.30 $ & $ -	0.24  $ \\
8   & $ 10.574 $ & $ +0.31 $ & $ -	0.25  $ \\
9   & $ 11.815 $ & $ +0.33 $ & $ -	0.27  $ \\
10  & $ 13.029 $ & $ +0.33 $ & $ -	0.28  $ \\
11  & $ 14.249 $ & $ +0.34 $ & $ -	0.29  $ \\
12  & $ 15.437 $ & $ +0.34 $ & $ -	0.29  $ \\
\end{tabular} 
\end{center}

\tablenotetext{a} {The central value of the $\bar{k}_{\ell}$'s
is taken for the $s=1.0$ case. The last two columns are 
the deviations in $\bar{k}_{\ell}$ from the central value
when
$s=1.5$ and $0.5$  are used respectively}

\tablenum{1}
\caption{Variation of estimated arguments $\bar{k}_{\ell}$
with $s$. \label{tbl-1}}

\end{table*}

\clearpage

\figcaption{Kernel function $K_{\ell}(k)$ for $\ell = 2$ ({\it solid}),
$\ell = 9$ multiplied by 100 ({\it short dash}) and
$\ell = 18$ multiplied by 1000 ({\it long dash}) \label{fig1}}

\figcaption{Spectrum of PDF as derived 
from COBE's 2 yr angular power spectrum and 
best fit to a power-law function. \label{fig2}}

\figcaption{COBE 2 yr angular power spectrum
({\it points}) compared to the power spectrum 
corresponding to our best power-law fit of $P(k)$
in Fig. 2. The external lines define the $\pm 1 \sigma$
band expected from cosmic variance.
 \label{fig3}}

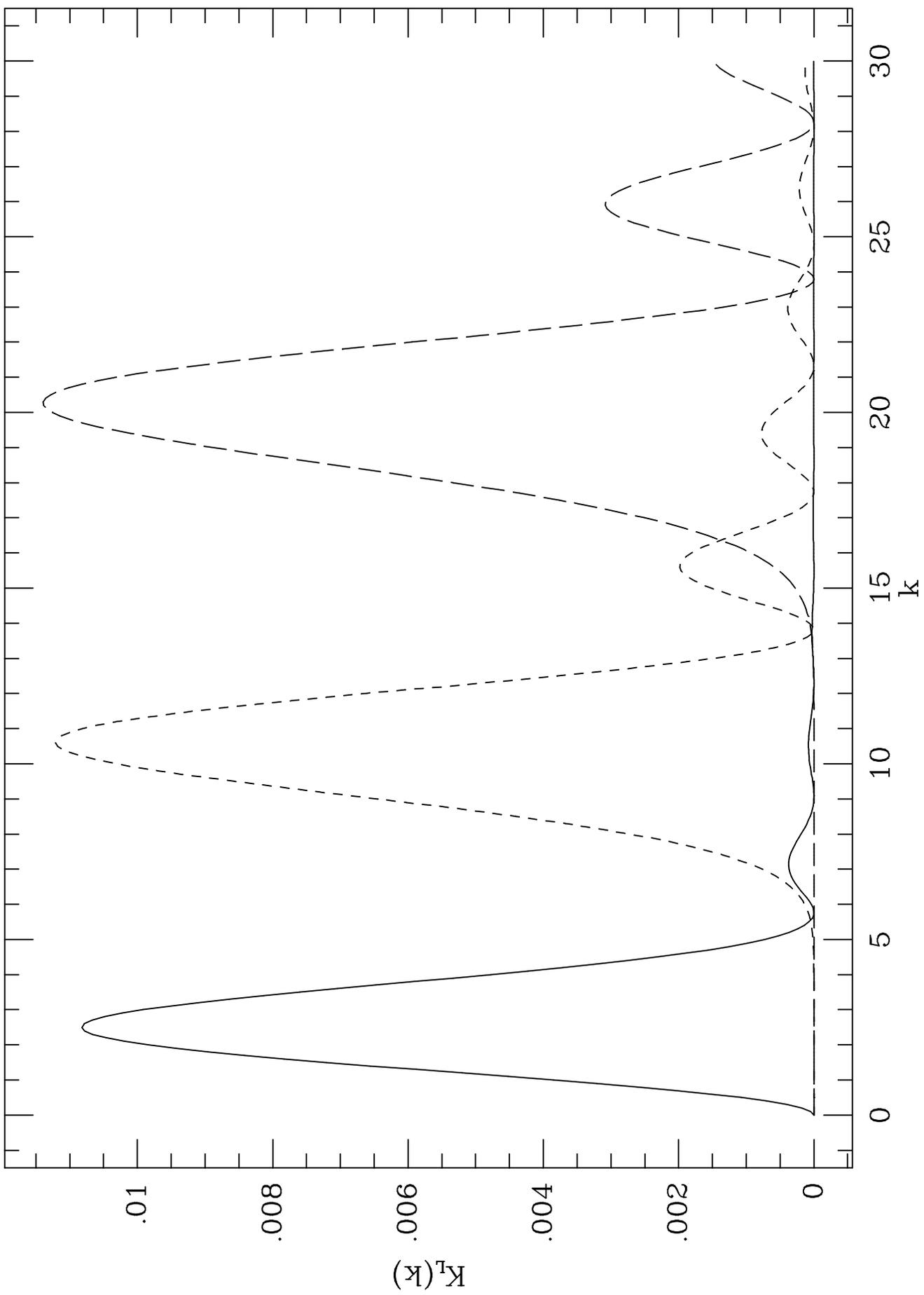
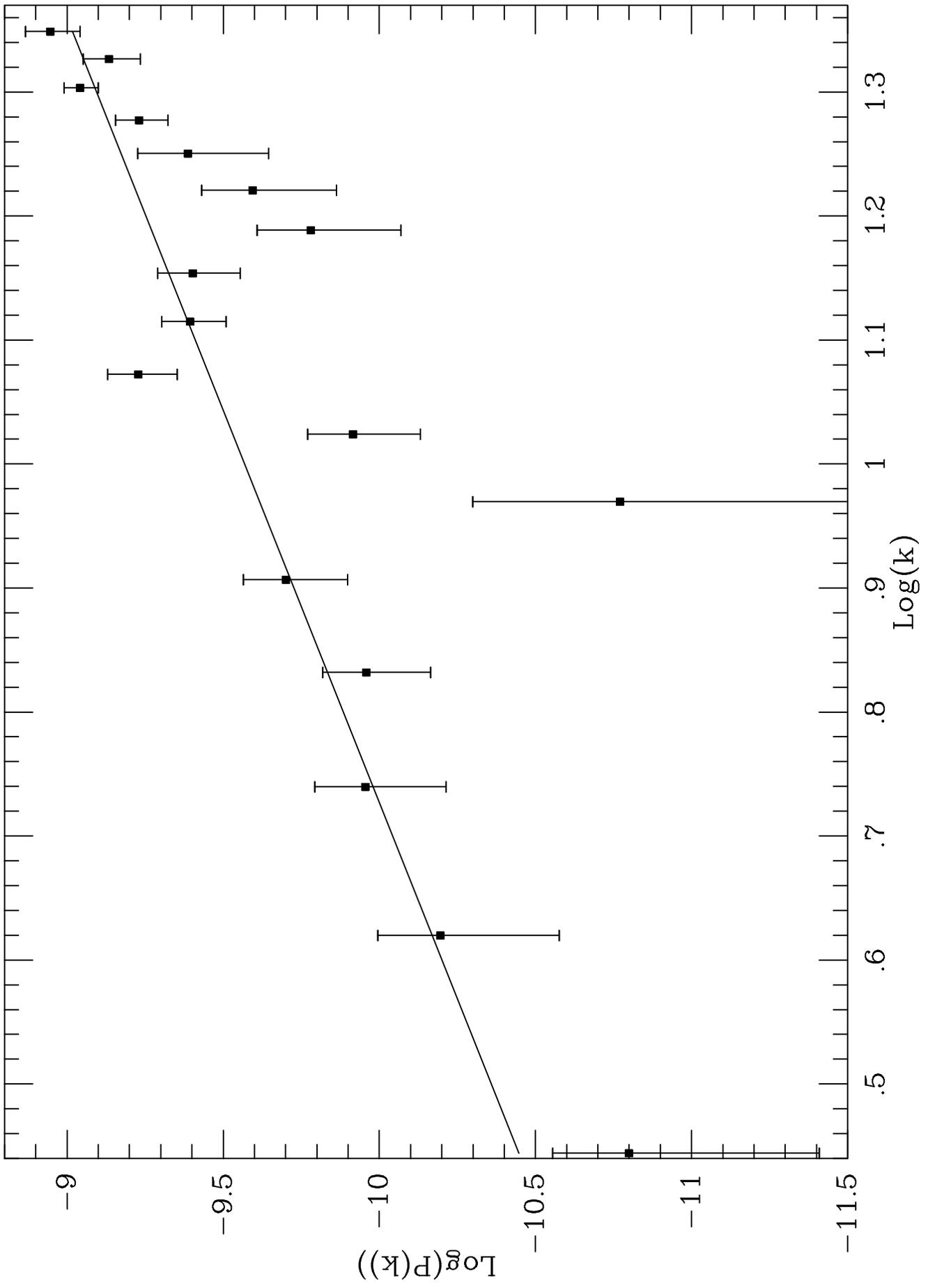
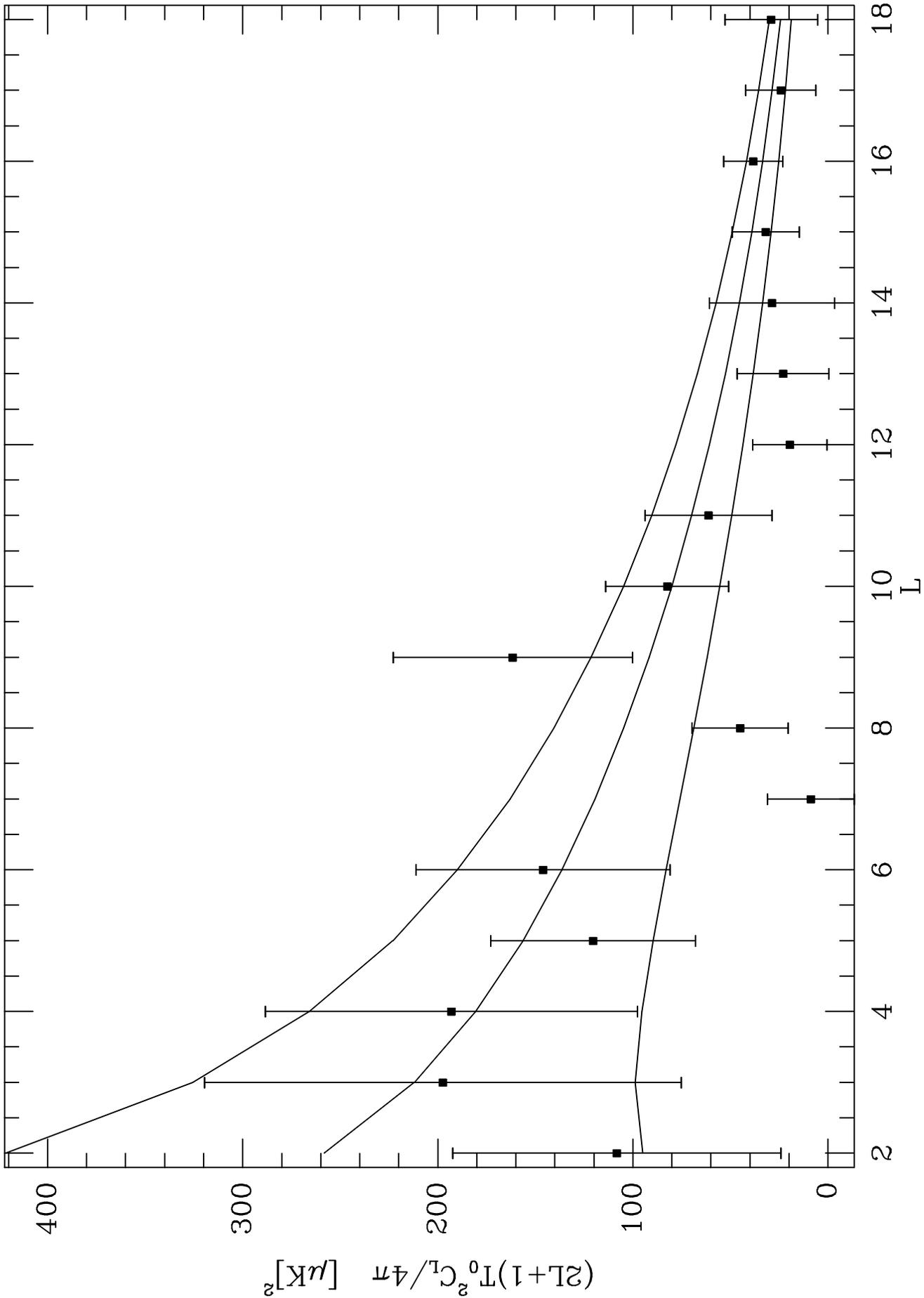
\end{document}